\documentclass[aps,prl,amsmath,amssymb,amsfonts,superscriptaddress,floatfix,twocolumn]{revtex4}
\usepackage{amsmath}
\usepackage{graphicx}
\newcommand{\be}{\begin{equation}}
\newcommand{\ee}{\end{equation}}
\begin{document}
\title{Inference from Matrix Products: A Heuristic Spin Glass Algorithm}
\author{M.~B.~Hastings}
\affiliation{Center for Nonlinear Studies and Theoretical Division,
Los Alamos National Laboratory, Los Alamos, NM, 87545}
\affiliation{Kavli Institute for Theoretical Physics, University of California, Santa Barbara, CA 93106}
\begin{abstract}
We present an algorithm for finding ground states of two dimensional
spin glass systems based on ideas from matrix product states in
quantum information theory.  The algorithm works directly at
zero temperature and defines an
approximate ``boundary Hamiltonian"
whose accuracy depends on a parameter $k$.  We test the algorithm against exact
methods on random field and random bond Ising models, and we find that
accurate results require a $k$ which scales roughly polynomially with
the system size.  
The algorithm also performs well when tested on small systems
with arbitrary interactions, where no fast, exact algorithms exist.
The time required is significantly less than
Monte Carlo schemes.
\end{abstract}
\maketitle

The problem of finding spin glass ground states is a major problem in
statistical physics.  Even for a two dimensional Ising system,
this problem is NP-hard if the bonds and fields are arbitrary\cite{nph}.
For this reason, exact algorithms will often be unable to find the correct ground state in a reasonable time, and
heuristic algorithms become necessary.

Monte Carlo algorithms, including parallel tempering\cite{pt},
extremal optimization\cite{eo}, and others, present one important
class of heuristic algorithms.  Their performance can be tested
against exact algorithms in two important cases, the case of planar
Ising models where all the fields vanish but the bonds are
arbitrary and the case of ferromagnetic
Ising models with arbitrary magnetic fields, where fast exact algorithms
are available.  These problems are the random bond and random
field problems, respectively.

Unfortunately, these Monte Carlo algorithms can still take exponentially
long time to find the true ground state even in these cases\cite{eo2,mcbench}.
In this paper,
we present an approach to finding spin glass ground states inspired
by applications of density matrix renormalization group (DMRG)\cite{dmrg}
to two dimensional classical
thermodynamics\cite{tmdmrg}.  We call our approach the inference matrix product (IMP) algorithm.
IMP, {\it working directly at zero temperature}, greatly
reduces
the time required compared to transfer matrix DMRG for these systems.

We consider the following Hamiltonian:
\be
E=\sum_{<i,j>} J_{ij} \sigma_i \sigma_j + h_i \sigma_i,
\ee
where $\sigma_i=\pm 1$.  The couplings $J_{ij}$ are between nearest neighbor
spins on a two dimensional lattice (we consider square lattices in this
paper, but the development is similar for other lattices).  We consider
a system of length $L$ and width $W$, with open boundary conditions, with
the coordinate in the length direction labeled from $1$ to $L$.

We begin by presenting the algorithm for this problem.  We then compare to exact
algorithms on certain cases, and discuss the possible role of criticality
in describing these systems.

{\it Inference Matrix Product (IMP) Algorithm---}
In this section, we present the IMP algorithm.  We begin by
some background, giving the solution of the problem in the case $W=1$.  In
other words, we begin with
a one dimensional chain.  Let $E^{(n)}(\sigma_1,...,\sigma_n)$ denote
the energy of the first $n$ spins for a subchain of length $n \leq L$.
Let 
\be
E_\leftarrow^{(n)}(\tau)=
{\rm min}_{\sigma_1,...,\sigma_{n-1}} E^{(n)}(\sigma_1,...,\sigma_{n-1},\tau),
\ee
denote the minimum energy of a subchain of length $n\leq L$ with the
last spin fixed to have value $\tau$.   The subscript
``$\leftarrow$" denotes that the energy is over a subchain to the {\it left} of the
spin we consider, and later we introduce energies $E_\rightarrow$;
these energies are boundary Hamiltonians describing the energy as
a function of spins at the boundary of the chain.
Then we have the recursion relations:
\begin{eqnarray}
\label{recrel}
E_\leftarrow^1(\tau)=h , \\ \nonumber
E_\leftarrow^{(n+1)}(\tau)=
{\rm min}_{\sigma} \Bigl( E_\leftarrow^{(n)}(\sigma)+J_{n,n+1} \sigma \tau + h_{n+1} \tau
\Bigr).
\end{eqnarray}
These recursion relations can be solved in time $O(L)$ to find
$E_\leftarrow^{(L)}(\tau)$, which can be minimized
over $\tau$ to find the ground
state energy of the chain.

Similarly, we can find the ground state as follows: by minimizing
$E_\leftarrow^{(L)}(\sigma_L)$, we can fix $\sigma_L$.  Then, we minimize
$E_\leftarrow^{(L-1)}(\sigma_{L-1})+J_{L-1,L} \sigma_{L-1} \sigma_L$ 
to fix $\sigma_{L-1}$.
Proceeding in this fashion moving back to the left along the chain,
we find the ground state.
The approach we have described is variously referred to as transfer
matrix, dynamic programming, or belief propagation in different fields.

To solve the problem with $W>1$, we define $E_\leftarrow^{(n)}(\tau_1,...,\tau_W)$
to denote the minimum energy of the subsystem with length $n$ and width $W$,
with the last column of spins fixed to be $\tau_1,...,\tau_W$.
We use coordinates $(r,c)$ to denote the row and column of the
spins, with $r=1,...,L$ and $c=1,...,W$.  Then,
\begin{eqnarray}
\label{rec2d}
E_\leftarrow^{(1)}(\tau_1,...,\tau_W)&=& C^{(1)}(\tau_1,...,\tau_W), \\ \nonumber
E_\leftarrow^{(n+1)}(\tau_1,...,\tau_W)&=&
{\rm min}_{\sigma_1,...,\sigma_W}\Big(
E_\leftarrow^{(n)}(\sigma_1,...,\sigma_W)\\ \nonumber &+&
\sum_{j=1}^{W} J_{(n,j),(n+1,j)} \sigma_j \tau_j+C^{(n+1)}(\tau_1,...,\tau_W)\Bigr),
\end{eqnarray}
where the energy of a column of spins $C^{(n)}$ is
$C^{(n)}(\tau_1,...,\tau_W)=
\sum_{j=1}^{W-1} J_{(n,j),(n,j+1)} \tau_j \tau_{j+1}
+\sum_{j=1}^W h_{(n,j)} \tau_j$.

Unfortunately, the brute force solution of Eqs.~(\ref{rec2d}) is not practical.
There are $2^W$ possible spin configurations on a given column,
and hence the time required scales as $O(L 2^{2W})$.

The approach we follow
is designed to solve this problem.  Our approach, inspired by the idea
of matrix product states, is to define an approximation to
$E_\leftarrow^{(n)}(\tau_1,...,\tau_W)$, which we denote
$\tilde E_{\leftarrow,k}^{(n)}(\tau_1,...,\tau_W)$ and which we refer to as
a matrix product energy.  
The quantity $k$ is an integer, which we call the bond dimension.  For
larger values of $k$ this approach becomes more accurate, but
at the same time more computationally costly.
This approximation will have the property
that
\be
\label{variational}
\tilde E_{\leftarrow,k}^{(n)}(\tau_1,...,\tau_W)
\geq
E_{\leftarrow}^{(n)}(\tau_1,...,\tau_W).
\ee

We define the energy $\tilde E_{\leftarrow,k}^{(n)}(\tau_1,...,\tau_W)$ by introducing
auxiliary variables, called bond variables, on each vertical bond in
the column of spins.  We label these bond variables
$\alpha_{i,i+1}$, for $i=0,...,W$, and each bond variables
may assume $k$ different values.  Note that although we consider open boundary
conditions for the interactions, we introduce bond variables
$\alpha_{0,1}$ and $\alpha_{W,W+1}$ connecting above the top spin and
below the bottom spin.  This is done for notational convenience later.

We now define an auxiliary problem which
is a one dimensional problem of the spins $\tau$ and the bond variables
$\alpha$ interacting along the column, with energy $F^{(n)}$,
and we define the energy
$\tilde E_{\leftarrow,k}^{(n)}(\tau_1,...,\tau_W)$ to be the minimum of $F^{(n)}$ over
the bond variables $\alpha$.  Specifically, we set
\be
F^{(n)}
\equiv
\sum_{i=1}^{W} F_i^{(n)}(\tau_i,\alpha_{i-1,i},\alpha_{i,i+1}),
\ee
where the functions $F_i$ are described below.  Thus, each spin $\tau$
is coupled to the bond variable above and below it.
We set $J_{0,1}=J_{W,W+1}=0$.

Note that
the energy $E_\leftarrow^{(1)}(\tau_1,...,\tau_W)$ from the first line of
Eq.~(\ref{rec2d}) can be described exactly by a matrix product energy
with $k=2$ as follows.  To see this, label the two states of the bond
variable by $\pm 1$, just as we label the states of the spin by $\pm 1$.
Then, set
$F_i^{(1)}(\tau_i,\alpha_{i-1,i},\alpha_{i,i+1})=h_i \tau_i+
J_{(1,i-1),(1,i)}\alpha_{i-1,i}\tau_i$ if
$\tau_i=\alpha_{i,i+1}$ and
$F_i^{(1)}(\tau_i,\alpha_{i-1,i},\alpha_{i,i+1})=\infty$ otherwise.

It is convenient for numerical purposes to ``encode" the spins in the bond
variables; that is, we will consider only states such that given
$\alpha_{i,i+1}$, there is only one choice of $\tau_i$ which gives
a finite energy.  Then, given a matrix product energy, $\tilde E^{(n)}_{\leftarrow,k}(\tau_1,...,\tau_W)$
with bond dimension $k$, we can find a matrix product energy 
$\tilde E^{(n+1)}_{\leftarrow,2k}(\tau_1,...,\tau_W)$ which has bond dimension $2k$
and which equals 
${\rm min}_{\sigma_1,...,\sigma_W}(
\tilde E_{\leftarrow,k}^{(n)}(\sigma_1,...,\sigma_W)+
\sum_{j=1}^{W} J_{(n,j),(n+1,j)} \sigma_j \tau_j+C^{(n+1)}(\tau_1,...,\tau_W)\Bigr)$.
We refer to this as ``propagating" the energy to the right.

As this propagation proceeds, $k$ increases exponentially.  Thus, an
exact way to compute the ground state energy is to proceed as follows:
initialize the energies $F_{i}^{(1)}$ to give a matrix product energy
with $k=2$,
and propagate to the right, until the final matrix product energy has
bond dimension $k_W=2^L$.
Then, the task of finding the optimum spin configuration
proceeds as follows.  First, compute the optimum spin and bond variable
configuration
for the one dimensional problem defined by energy $F^{(L)}$; since this
is a one dimensional problem, the minimization can be readily done in
a time $O(k_W^2 W)$ by the techniques described above for one dimensional
chains.  Then, fix the spins on column $L$ to the values which minimize
$F^{(L)}$, and compute the optimum spin and bond variable configuration for
the problem $F^{(L-1)}+\sum_{i=1}^W J_{(L-1,i),(L,i)} \sigma_{(L-1,i)}
\sigma_{(L,i)}$.  Proceeding in this fashion, we find the ground state.

Of course, this approach is also exponentially costly.  Therefore, we
define below a removal of redundances and a
``truncation".  We fix a maximum bond dimension $k_{max}$,
and when the bond dimension of a matrix product energy 
$\tilde E^{(n)}_{\leftarrow,k}$
exceeds
$k_{max}$ we truncate by finding a good approximation 
$\tilde E^{(n)}_{\leftarrow,k_{max}}$.
This truncation is achieved by choosing, for
each bond, a subset of size $k_{max}$
of the bond variables, and only allowing the bond variables to assume
those particular choices.  Then, we will have the property that
for $k\geq k_{max}$,
\be
\tilde E^{(n)}_{\leftarrow,k_{max}}\geq \tilde E^{(n)}_{\leftarrow,k}.
\ee
The way we choose this subset is described below.  Unsurprisingly,
the best choice of the bond variables will depend on the optimum configuration
of the spins to the right of those we have considered thus far, and
at this point we have no knowledge of the best configuration of those
spins.  Therefore,
after propagating $\tilde E_{\leftarrow,k}$ to the right and truncating,
we then propagate back $\tilde E_{\rightarrow,k}$ to the left so that on the
propagation back to the left we will use the
$\tilde E^{(n-1)}_{\leftarrow,k_{max}}$ which we have previously computed to
decide the best truncation of $\tilde E^{(n)}_{\rightarrow,k}$.  

The IMP algorithm is defined as follows:
\begin{itemize}
\item[1.] Initialize the energies $F_i^{(1)}$
and set the bond dimension for the first column, $k_1$, equal
to $2$.  Set $n=1$.

\item[2.] Propagate to the right to compute $\tilde E^{(n+1)}_{\leftarrow,k}$ from 
$\tilde E^{(n)}_{\leftarrow,k}$, and set the bond dimension $k_{n+1}$ for the $n+1$-st column
equal to $2 k_n$,.

\item[3.] If $k_{n+1}\geq k_{max}$, for some $k_{max}$, remove redundancies
and truncate as described
below.

\item[4.] Increment $n$ and if $n<W$ goto step $2$.

\item[5.] Repeat steps $1$ to $4$, but this time
propagate $\tilde E^{(n)}_{\rightarrow,k}$ from right to left.

\item[6.] Compute the optimum spin configuration, working from left to right.
\end{itemize}

{\it Removing Redundancies---}
Before truncating, we remove redundancies.  In practice, this step
does not have much effect on performance,
and so on first reading the reader
should skip down to the section on truncation.
The bond variable
$\alpha_{0,1}$,
leaving the top spin connects to only one spin, $\tau_1$.  Therefore, we can
replace any matrix product energy in which this bond variable can assume
$k$ different variables by one in which the bond variable can assume only
a single value, which we set to zero, by replacing
$F_1(\tau_1,\alpha_{0,1},\alpha_{1,2})$ by $F'$ with
$F'(\tau_1,0,\alpha_{1,2})={\rm min}_{\alpha_{0,1}}
F_1(\tau_1,\alpha_{0,1},\alpha_{1,2})$.
We can do something similar for the bottom variable, $\alpha_{W,W+1}$,
except since this variable encodes the state of $\tau_W$, we must allow
it to assume two different values.

For all the bond variables in between, we can remove redundant states as
follows.  Consider a bond variable $\alpha_{i,i+1}$.
If there are two states of this bond variable, $l_1$ and $l_2$,
which have the property that
$F_i(\tau_i,\alpha_{i-1,i},l_1)=F_i(\tau_i,\alpha_{i-1,i},l_2)$ for
all $\alpha_{i-1,i}$, then we replace $F_{i+1}$ by $F'$ with
$F'(\tau_{i+1},\alpha_{i,i+1},\alpha_{i+1,i+2})=
F_{i+1}(\tau_{i+1},\alpha_{i,i+1},\alpha_{i+1,i+2})$ if
$\alpha_{i,i+1}\neq
l_1$ and
$F'(\tau_{i+1},l_1,\alpha_{i+1,i+2}=
{\rm min}\Bigl(F_{i+1}(\tau_{i+1},l_1,\alpha_{i+1,i+2}),F_{i+1}(\tau_{i+1},l_2,\alpha_{i+1,i+2})\Bigr)$.
Now $\alpha_{i,i+1}$ is only allowed to assume $k-1$ different
values and no longer may assume the value $l_2$.  We can do a similar
procedure if
$F_{i+1}(\tau_{i+1},l_1,\alpha_{i,i+1})=F_{i+1}(\tau_{i+1},l_2,\alpha_{i,i+1})$.

Finally, if it turns out to be the case for a bond variable
$\alpha_{i,i+1}$ we find that
$F_i(\tau_i,\alpha_{i-1,i},l_1)<F_i(\tau_i,\alpha_{i-1,i},l_2)$ for all
$\alpha_{i-1,i}$ and all $\tau_i$, and that
$F_{i+1}(\tau_{i+1},l_1,\alpha_{i,i+1})<F_{i+1}(\tau_{i+1},l_2,\alpha_{i,i+1})$
for all $\alpha_{i,i+1}$ and all $\tau_{i+1}$, then we can remove the choice
$l_2$.

{\it Truncation and Parameters---}
We now describe the truncation procedure.  We describe the procedure
in the case of the propagation back to the left (step 5), where we truncate
a matrix product energy $E^{(n)}_{\rightarrow,k}$.
To help guide the choice of truncation, we use the matrix product energy
$E^{(n-1)}_{\leftarrow,k}$.  We will write
$F_{\rightarrow}$ and $F_{\leftarrow}$ to refer to the auxiliary
problems used to define
$E^{(n)}_{\rightarrow,k}$ and
$E^{(n-1}_{\leftarrow,k}$.
Further, we write $J_i$ to refer to
$J_{(n-1,i),(n,i)}$, and we use $\sigma^r_i$ to refer to spins in column
$n$ and $\sigma^l_i$ to refer to spins in column $n-1$, and use
$\alpha_{i,i+1}$ to refer to bond variables in $F_\rightarrow$ and
$\beta_{i,i+1}$ to refer to bond variables in $F_\leftarrow$.

We begin by truncating bond variable $\alpha_{W,W+1}$.  We consider the
energy
\be
\label{etdef}
E_{tot}=
F_\rightarrow+F_\leftarrow+\sum_i J_i \sigma^l_i \sigma^r_i.
\ee
For each of the $k$ different choices of $\alpha_{W,W+1}$, we minimize
this energy over the other bond variables $\alpha$ and $\beta$ and the spins.
We then truncate by only keeping the $k_{max}$ values of $\alpha_{W,W+1}$
which lead to the lowest energy.  We then truncate the bond variable
$\alpha_{W-1,W}$ by computing, for each of the $k$ different choices of
this bond variable, the minimum of the energy over all the other
the bond variables and spins, and again keep only the $k_{max}$ values which
minimize the energy.  We proceed in this way until we have truncated all
the bond variables.  Note that in, for example, the truncation of
$\alpha_{W-1,W}$, when we minimize the energy we only allow bond variables
$\alpha_{W,W+1}$ to assume one of the $k_{max}$ different states we kept
on the first truncation step.

To do this truncation requires computing the energy the minimum energy of
a one dimensional system, $E_{tot}$.  This can be done by computing
the minimum energy of a subchain consisting of the bottom $h$ bonds, where
the top two bonds are constrained to have values $\alpha$ and $\beta$.
Propagating this energy, just as in Eq.~(\ref{recrel}) gives the desired
answer.  The most naive algorithm would take a time $O(W k^4)$ to do the
truncation.  However, we can take advantage of the simple form of the
interaction between the two chains, and do the entire truncation
step in a time $O(W k^3)$, by propagating the energy upward for each of
the four choices of the top two spins.

In step (3), we perform the truncation similarly to here, except that
we set $E_{tot}=F_\leftarrow$, without considering the interaction between two
columns of spins.  While this is much faster, taking only a time $O(W k^2)$,
the propagation back in the other direction in step (5) is necessary in
many cases to obtain accurate results.  The total time for the algorithm
then scales as $O(L W k^3)$.

{\it Results on Random Bond and Random Field Ising Models---}
We compared the performance of this algorithm to exact solutions on the
random bond and random field Ising models.  
In the random bond case, 
we chose each bond strength independently at random
from a uniform distribution between $-1$ and $1$.  All the magnetic fields
were set equal to zero.  While polynomial algorithms exist for
planar Ising models with no fields, as in this case\cite{rbim}, but
we used the spin glass server to find exact solutions\cite{sgserver} for
problems of size $L=W=5,10,20,40$.  In Fig.~1(a) we show the performance of
the algorithm as a function of $k$ for these sample sizes.  As a
rough estimate, the bond dimension for the random bond problem
to obtain $50\%$ accuracy
scales as $L^{4/3}$, so the total time computational
time required scales roughly as $L^6$.
More numerical work is needed to truly ascertain the scaling of $k$ with
$L$, but certainly we do not need exponentially large values of
$L$ to obtain good accuracy.

\begin{figure}
\centerline{
\includegraphics[scale=0.3,angle=270]{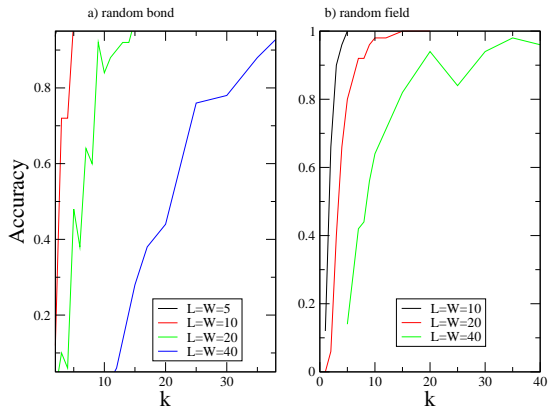}}
\caption{Fraction of correctly identified ground states as a function
of $k$ for random bond (a) and random field (b) problems (50 realizations
in each case).  The curve
with $k=5$ is barely visible in $a$ since perfect accuracy out
of 50 samples was obtained with $k=3$.  A small number (50) of samples was
deliberately chosen to illustrate that for some individual samples increasing
$k$ can occasionally lead to a loss of accuracy.}
\vspace{5mm}
\end{figure}

In the random bond case, we set all bonds ferromagnetic
with unit strength, and chose the fields independently at random from
a uniform distribution between $-H$ and $+H$ for some parameter $H$.
For small values of $H$, we tend to produce only a single
domain for small systems, and so finding the ground state is simple (however,
for any fixed $H$, the ground state will acquire domains for large $L$\cite{rfim}).
For large values of $H$, each spin aligns
with the local field and again finding the ground state is simple.
The hardest cases seem to arise for moderate values of $H$.
In Fig.~1(b), we show simulations of accuracy
as a function of $k$, so systems sizes $L=W=10,20,40$ with $H=5$.  This
value of $H$ was chosen to provide a system with large domains.  Exact
solutions were performed by an open source push-relabel algorithm\cite{prf}.

Finally we tested on systems with arbitrary fields and bonds.  Exact solutions
were available for small system sizes by brute force; for larger system
sizes we can use the IMP algorithm for large $k$ to test the algorithm
for small $k$.  The accuracy of the algorithm as a function of $k$ is
comparable in this case to the random bond and field problems, as far
as we were able to test it.

{\it Discussion---}
We have presented an algorithm for finding ground states of
two dimensional glassy systems.  The sweep back left is
important for accuracy.
Working at zero temperature
is important for speed;
doing transfer matrix DMRG\cite{tmdmrg} or
TEBD\cite{ov} for a system at non-zero temperature
would require a time $O(k^3)$ to do the initial sweep to the
right, and would require a time $O(k^5)$ in the sweep back left,
compared to $O(k^2)$ and
$O(k^3)$ respectively for IMP.
Note that the sweep back left to improve truncation was not
considered in those algorithms since they
were not applied to glassy systems.
Running for
$k=20$ with $L=W=40$ requires roughly 3 seconds of CPU time on a 2.0 Ghz
G5, much faster than Monte Carlo.
Given that for arbitrary fields and bonds the problem is NP-hard,
the required $k$ for an arbitrary problem may scale exponentially
with $L$.  However, for typical problems, generated randomly, a polynomial
scaling of $k$ is quite possible.  This may be related
to possible conformal invariance\cite{confin1,confin2} in
the random bond problem, given the relationship between
entanglement entropy and the scaling of $k$ for
one dimensional quantum systems\cite{poly,renyi,renyi2}.

It would be interesting to extend to three dimensional
problems, by defining a boundary Hamiltonian on 
a plane, rather than a column.
Truncating the bond variables would require minimizing
the energy of a two 
dimensional problem, which can be done using the techniques described
here.  This idea of descending to one dimension lower
is reminiscent of
\cite{descent}.  Another possible extension would be to combine the
algorithm with Monte Carlo methods, by using randomness in the
choice of states
to keep in the truncation,
and doing several sweeps until the ground state is
found.

{\it Acknowledgments---}
The spin glass server was used to perform certain
simulations.  This work was supported by
U. S. DOE Contract No. DE-AC52-06NA25396.

\end{document}